\documentclass[11pt,a4paper]{article}
\usepackage{jheppub}

\def\beq{\begin{equation}}
\def\eeq{\end{equation}}

\usepackage{amsmath}
\usepackage{epsfig,epsf}
\usepackage{graphicx}
\usepackage{verbatim}

\title{Conformal symmetry of JIMWLK Evolution at NLO.}
\author[a]{Alex Kovner,}
\author[b]{Michael Lublinsky,} 
\author[b]{Yair Mulian}

\affiliation[a]{Physics Department, University of Connecticut, 2152 Hillside road, Storrs, CT 06269, USA}
\affiliation[b]{Physics Department, Ben-Gurion University of the Negev, Beer Sheva 84105, Israel}
 
\date{\today}
\abstract{ 
 We construct the  Next to Leading Order JIMWLK Hamiltonian for high energy evolution in ${\cal N}=4$ SUSY theory, and show that it possesses conformal invariance, even though it is derived using sharp cutoff on rapidity variable.  The conformal transformation properties of Wilson lines are not quite the naive ones, but at NLO acquire an additional anomalous piece. We construct explicitly the inversion symmetry generator. We also show how to construct for every operator $O$, including the Hamiltonian itself, its "conformal extension" $\cal O$, such that it transforms under  the inversion in the naive way. 
}

\date{\today}
\begin{document}
\maketitle

\pagestyle{empty}
\newpage

\mbox{}

\pagestyle{plain}

\setcounter{page}{1}




\date{\today}

\section{Introduction}
In recent years a lot of attention has been devoted to development and phenomenological applications of the theory of perturbative saturation \cite{GLR}. The main physical idea of this approach is that at high enough energy hadronic wave function resembles a dense gluon cloud, sometimes referred to as the Color Glass Condensate (CGC) \cite{cgc}. When the energy is high enough the density scale becomes large and the physics becomes essentially perturbative and tractable. 

The theoretical description of the energy evolution of the wave function towards such a dense state at leading order in $\alpha_s$ has been long known. It is given by the so called JIMWLK equation \cite{jimwlk}, or equivalently Balitsky hierarchy \cite{Bal}. It generalizes the well known BFKL equation \cite{bfkl},\cite{bkp} by including finite density effects in the hadronic wave function. 

The mean field approximation to the JIMWLK equation, the so called Balitsky-Kovchegov (BK) equation \cite{Bal},\cite{KOV} has been used extensively in the last several years in phenomenological applications, that include fits to DIS low x data \cite{lublinsky},\cite{AAMQS} as well as various aspects of p-p and p-A data\cite{review}.
For phenomenological applications it is crucial to include perturbative corrections beyond leading order, since they are known to lead to large effects already in the linear BFKL framework \cite{NLOBFKL}. Currently only the corrections due to running coupling constant  (\cite{BalNLO},\cite{weigertrun}) are included in the numerical work,
although there has been recent progress in understanding of the more problematic gluon emission contributions \cite{beuf}. 

Significant progress in including the full set of next to leading order corrections in the high energy evolution was achieved by Balitsky and Chirilli \cite{BC}. This work presented the complete set of NLO corrections to the evolution of the scattering amplitude of the fundamental dipole in QCD. Subsequently analogous calculation was performed in the ${\cal N}=4$ super Yang Mills theory \cite{N=4}. Recently Grabovsky \cite{Grab} computed certain  parts of the NLO evolution equation for three-quark 
singlet amplitude in the $SU(3)$ theory.

Using the results of \cite{BC} and \cite{Grab} in a recent paper \cite{nlojimwlk} we have derived the complete operator form of the JIMWLK Hamiltonian at next to leading order. This paper appeared simultaneously with \cite{BClast}, which directly calculated most elements of the general Balitsky hierarchy at NLO. We also note that similar results have been independently obtained in \cite{Simon}. For the sake of self-completeness of the present paper,
 in Appendix A we quote the main result of  ref. \cite{nlojimwlk}.  In Appendix B
we provide some insight on how the result is obtained, while  a more detailed report of our derivation and a  comparison with \cite{BClast} will appear 
in a forthcoming separate publication \cite{KLMatwork}.

Even though the JIMWLK NLO Hamiltonian is now available, there are theoretical questions about it that still have to be addressed. In this paper we address one such question, namely that of conformal invariance. The leading order JIMWLK equation is conformally invariant when applied on gauge invariant states. 
This holds even in QCD, although conformal invariance of the classical Yang-Mills action is violated by the quantum anomaly, since the coupling constant renormalization is necessary only beyond leading order. 

The NLO evolution of the color dipole derived in \cite{BC} as well as the explicit form of the NLO JIMWLK Hamiltonian given in \cite{nlojimwlk} is not conformally invariant. There are two sources for the violation of conformal symmetry at this level. One is due to the genuine quantum anomaly associated with the introduction of renormalization scale. The other one is due to the fact that the calculations involve hard cutoff in rapidity space, which itself is not conformally invariant \cite{FF}. In principle it should be possible to eliminate this latter source of noninvariance by employing an explicitly conformally invariant rapidity cutoff. However, it is not known how to do it explicitly. Instead, it was shown in \cite{N=4} that in the particular case of the dipole evolution it is possible to redefine the dipole operator in such a way that its evolution becomes conformally invariant, up to the running coupling effects. In this paper we show that the reason for it, is that conformal invariance is in fact present in the NLO JIMLWK Hamiltonian, albeit the conformal transformation of the Wilson lines is slightly different from the naive one. To avoid dealing with the running coupling effects, we will consider here the ${\cal N}=4$ super Yang-Mills theory. On the level of JIMWLK equation this theory is very similar to QCD. However it has no conformal anomaly and is conformally invariant as a full Quantum Field Theory.

We show that the NLO JIMLWK equation for ${\cal N}=4$ theory has exact conformal invariance, even though it is derived with sharp rapidity cutoff. The conformal transformation of the Wilson line operators is perturbatively different from the naive one, and this is the origin of apparent noninvariance of the $H^{NLO\ JIMWLK}$ as well as the dipole evolution equation. The modified transformation is an exact symmetry of the Hamiltonian\footnote{To be more precise, within the perturbative NLO framework the Hamiltonian $H^{NLO\ JIMWLK}$ is invariant up to terms of order $O(\alpha_s^2)$.}. Our functional formalism  is not only an explicit realization of the
idea that evolution kernels could be modified  as discussed in \cite{FF,N=4}, but a proof that such modification is possible for any operator in the theory\footnote{Results, in many respects similar to ours were obtained independently in \cite{Simon}.  We thank Simon Caron-Huot for sharing his results prior to publicaiton.}

We show how the properties of the generator of the conformal transformation allow one to define operators with naive conformal transformation properties. Evolution equations satisfied by these ``conformal operators'' are invariant under naive conformal transformation. For the color dipole operator this procedure results in the same definition of the conformal dipole as given in \cite{N=4}.
Applying this general procedure to the baryon operator in $SU(3)$ theory, we derive an expression for the  "conformal baryon". We also derive the operator form of the Hamiltonian, which generates evolution of the ``conformal operators''. This Hamiltonian itself is invariant under naive conformal transformation.

We present the NLO JIMWLK Hamiltonian for the ${\cal N}=4$ SUSY in  the next  Section (2). Calculational details relevant for this derivation can be found in  Appendix B.
In Section 3 we apply the inversion transformation to the Hamiltonian and compute the ``anomalous'' term responsible for apparent breaking of the conformal invariance.
 In Section 4 we modify the symmetry generator
and demonstrate that  the full JIMWLK Hamiltonian is indeed invariant under the new symmetry transformation. Appendix C is a supplement to this section.
Conformal operators, including conformally invariant Hamiltonian are presented in Section 5.  Section 6 contains a summary of the results.

\section{The NLO JIMWLK Hamiltonian.}
The JIMWLK Hamiltonian \cite{jimwlk} is the limit of the QCD Reggeon Field Theory (RFT),
applicable for computations of high energy  scattering amplitudes   of dilute (small parton number) projectiles on dense (nuclei) targets.  In general it predicts the rapidity evolution of any hadronic observable $O$ via the functional equation of the form
\beq\label{1}
\frac{d}{dY}\,{ O} \,=\,-\,H^{JIMWLK}\,{ O}
\eeq

The JIMWLK Hamiltonian defines a two-dimensional non-local field theory of a  unitary matrix (Wilson line) $S(x)$ which, in the high energy eikonal approximation represents the scattering amplitude of a quark at the transverse coordinate $x$.  
The leading order  Hamiltonian is:
\beq\label{LO}
H^{LO\, JIMWLK}\ =\ -\ \frac{1}{ 2}\int d^2 z\,d^2 x\,d^2 y \ M(x,y,z)\,\hat h(x,y,z)
\eeq
with  $\hat h$ being the Hamiltonian density:
\beq\label{h}
\hat h\ \equiv\  \left[ J^a_L(x)\,J^a_L(y)\ +\ J_R^a(x)\,J_R^a(y)\ -\ 2J_L^a(x)\,S^{ab}_A(z)\,J^b_R(y)\right]\,.
\eeq
Here $S_A$ is a unitary matrix in the adjoint representation - the gluon scattering amplitude.
The left and right $SU(N)$ rotation generators, when acting on functions of $S$ have the representation
\begin{eqnarray}\label{LR}
J^a_L(x)=tr\left[\frac{\delta}{\delta S^{T}_x}T^aS_x\right]-tr\left[\frac{\delta}{\delta S^{*}_x}S^\dagger_xT^a\right] ; \ \ 
J^a_R(x)=tr\left[\frac{\delta}{\delta S^{T}_x}S_xT^a\right] -tr\left[\frac{\delta}{\delta S^{*}_x}T^aS^\dagger_x\right]\,.
\end{eqnarray}
Here $T^a$ are $SU(N)$ generators in the fundamental representation. The leading order dipole kernel  is given by
\begin{equation}
M(x,y;z)\,=\,\frac{\alpha_s}{2\,\pi^2}\,\frac{(x-y)^2}{X^2\,Y^2}\,.
\end{equation}
We use the notations of ref. \cite{BC} $X\equiv x-z$,  $X^\prime\equiv x-z^\prime$, $Y\equiv y-z$,    $ Y^\prime \equiv y-z^\prime$,
$W\equiv w-z$,    $ W^\prime \equiv w-z^\prime$, and $Z\equiv z-z^\prime$. 
The dipole form of the kernel is appropriate when the Hamiltonian acts on gauge invariant observables. If one wishes to consider evolution of non-gauge invariant states 
and/or is interested in non-singlet exchanges, like the reggeized gluon, the appropriate kernel is the Weizsacker-Williams kernel
\beq M(x,y;z)\rightarrow K(x,y;z)=-\frac{\alpha_s}{\,\pi^2}\,\frac{X\cdot Y}{X^2\,Y^2}\,.\eeq
In this paper we are only interested in the evolution of the gauge invariant sector, as ultimately only this sector of the theory is physical. Another reason to concentrate on the dipole form of the kernel is that even at leading order only the Hamiltonian with dipole kernel is explicitly conformally invariant.

At next to leading order the Hamiltonian contains terms with at most two factors of the adjoint Wilson line $S_A$, and at most three factors of the color charge density, since at this order at most two soft gluons are emitted in each step in the evolution.
More constraints on the form of the Hamiltonian come from the symmetries of the theory.
As discussed in detail in \cite{reggeon}, the theory must have $SU_L(N)\times SU_R(N)$ symmetry, which in QCD terms is the gauge symmetry of $|in\rangle$ and 
$|out\langle$ states and two discrete symmetries: the charge conjugation $S(x)\rightarrow S^*(x)$, and another $Z_2$ symmetry:  $S(x)\rightarrow S^\dagger(x)$,  $J^a_L(x)\leftrightarrow -J^a_R(x)$ which in \cite{reggeon} was identified with signature, and can be understood as the combination of charge conjugation and time reversal symmetry \cite{iancutri}.

Taking these constraints into account, the Hamiltonian can quite generally be written in terms of six kernels
\begin{eqnarray}\label{NLO}
&&H^{NLO\ JIMWLK}= \int_{x,y}K_{2,0}(x,y) \,\left[ J^a_L(x)\,J^a_L(y)\,+\,J_R^a(x)\,J_R^a(y)\right]\,\nonumber \\ 
&&-\,2\int_{x,y,z}\,K_{2,1}(x,y,z)\,J_L^a(x)\,S_A^{ab}(z)\,J^b_R(y)\nonumber \\
 &&+\int_{x,y,z,z^\prime} \,K_{2,2}(x,y;z,z^\prime)\,\left[f^{abc}\,f^{def}\,J_L^a(x)\, S^{be}_A(z)\,S^{cf}_A(z^\prime)\,J_R^d(y)\,-\,N_c\,J_L^a(x)\,S^{ab}_A(z)\,J^b_R(y)\right] \nonumber \\
 &&+\int_{w,x,y, z,z^\prime}K_{3,2}(w;x,y;z,z^\prime)f^{acb}\,\Big[J_L^d(x)\, J_L^e(y)\, S^{dc}_A(z)\,S^{eb}_A(z^\prime)\,J_R^a(w)
 \nonumber \\ &&  \ \ \ \ \ \ \ \ \ \ \   \ \ \ \ \ \ \ \ \ \ \  \ \ \ \ \ \ \ \ \ \ \ \   \ \ \ \ \ \ \ \ \ \ \   \ \ \ \ \ \ \ \ \ \ \  \ \ \ \ \ \ \ \ \ \ \ \  -\
J_L^a(w)\,S^{cd}_A(z)\,S^{be}_A (z^\prime)\,J_R^d(x)\,J_R^e(y)\,\Big]\nonumber \\
 &&+\int_{w,x,y, z}\, K_{3,1}(w;x,y;z)\,f^{bde}\,\Big[  J_L^d(x) \,J_L^e(y) \,S^{ba}_A(z)\,J_R^a(w)\,-\, 
J_L^a(w)\,S^{ab}_A(z)\,J_R^d(x)\,J_R^e(y) \Big]\, \nonumber \\
 &&+\int_{w,x,y}K_{3,0}(w,x,y)f^{bde}\left[\,J_L^d(x) \,J_L^e(y) \,J_L^b(w)\,-\, 
J_R^d(x)\,J_R^e(y)\,J_R^b(w)\right]\,.
\end{eqnarray}
All color charge density operators $J_{L(R)}$ in  (\ref{NLO}) are understood as placed to the right of all factors of $S$, and thus not acting on $S$ in the Hamiltonian. 

To determine the kernels in (\ref{NLO}) we calculate the action of the Hamiltonian on the color dipole and compare the result with the result of \cite{N=4}. 
The action of all terms in the Hamiltonian on a color dipole is given in Appendix B. Additionally we use the results of \cite{Grab} for the connected pieces of the evolution of the baryon operator. This corresponds to action of the Hamiltonian on the baryon operator $B(u,v,w)=\epsilon_{ijk}\epsilon^{lmn}S^{il}(u)S^{jm}(v)S^{kn}(w)$ and keeping only terms with at least one Wilson line and three color charge density operators, where no two operators $J_{L(R)}(x)$ act on the same coordinate of the baryon operator $B(u,v,w)$. The result of application of $H^{JIMWLK}$ on B gives directly the kernels $K_{3,2}$ and $K_{3,1}$, and here we do not present this calculation in any detail
(to appear in \cite{KLMatwork}). Comparison with \cite{Grab} gives 
\begin{eqnarray}\label{k}
 &&K_{3,2}(w;x,y;z,z^\prime)= \nonumber \\
 &&~~~~~~~~~ ~~~~~~~=\frac{i}{ 2}\Big[ M_{x,y,z}M_{y,z,z^\prime}+M_{x,w,z}M_{y,w,z^\prime} - M_{y,w,z^\prime} M_{x,z^\prime,z}-M_{x,w,z}M_{y,z,z^\prime}     \Big]\ln\frac{W^2}{W^{\prime\,2}}\nonumber\\ \nonumber \\
  &&K_{3,1}(w;x,y;z)=
  \int_{z^\prime}\, \left[K_{3,2}(y;w,x;z,z^\prime)-K_{3,2}(x;w,y;z,z^\prime)\right]
\end{eqnarray}
Note that we present these kernels here in a somewhat different form than in \cite{nlojimwlk}. The difference between eq.(\ref{k}) and similar expressions in \cite{nlojimwlk} are  terms which do not depend on one of the three coordinates $x,y$ or $w$. In the Hamiltonian, this amounts to additional operators with three $J$'s , which contain explicit factors of the type $Q^a_{L(R)}=\int d^2u \ J^a_{L(R)}(u)$. When such a factor appears in the rightmost position in the operator, the operator vanishes when acting on gauge invariant states, since such a state is annihilated by $Q^a_{L(R)}$. When $Q^a$ is not at the rightmost position, it can be commuted all the way to the right and then dropped. The commutator, which remains and cannot be neglected, involves one less power of $J$. Thus our choice of kernels $K_{3,2}$ and $K_{3,1}$ will be reflected by a somewhat different expression for the kernels $K_{2,1}$ and $K_{2,2}$ relative to those given in \cite{nlojimwlk}.

Comparing the result of the action of the Hamiltonian (\ref{NLO}) on a dipole with the dipole evolution calculated in \cite{N=4} we get the following relations:
\begin{eqnarray}
&&K_{2,2}(x,y;z,z^\prime) \ =\ \frac{\alpha_s^2}{16\,\pi^4}
\Bigg[\frac{(x-y)^2}{ X^2Y'^2(z-z^\prime)^2}\Big(1+\frac{(x-y)^2(z-z')^2}{X^2Y'^2-X'^2Y^2}\Big)-\nonumber \\
&&~~~~~~~~~~~~~~~~~~~~~~~~~~~~~~-\frac{(x-y)^2}{ X'^2Y^2(z-z^\prime)^2}\Big(1+\frac{(x-y)^2(z-z')^2}{X'^2Y^2-X^2Y'^2}\Big)\Bigg]\ \ln\frac{X^2{Y'}^2}{ {X'}^2Y^2}\\
&&K_{2,1}(x,y;z) =\frac{\alpha_s^2N_c}{16 \pi^3}\frac{(x-y)^2}{X^2 Y^2}\Big[\frac{\pi^2}{3}+2 \ln\frac{Y^2}{ (x-y)^2}\,\ln\frac{X^2}{ (x-y)^2}\Big]+
\nonumber \\ &&~~~~~~~~~~~~~~~~~~~~~~~~~~~~~~~~~~~~+\frac{i}{2}N_c\Big[K_{3,1}(y;x,y;z)+K_{3,1}(x;y,x;z)\Big] 
\end{eqnarray}
\begin{eqnarray}\label{k20}
\frac{4}{N_c}K_{2,0}(x,y)\,-\,i\,\Big\{K_{3,0}(y,x,y)+K_{3,0}(x,y,x)-K_{3,0}(y,y,x)-K_{3,0}(x,x,y)\ +\nonumber\\
+K_{3,0}(y,x,x)+K_{3,0}(x,y,y)\Big\}=\frac{\alpha^2}{4\pi^3}\int_z\frac{(x-y)^2}{X^2 Y^2}\Big[\frac{\pi^2}{3}+2 \ln\frac{Y^2}{ (x-y)^2}\,\ln\frac{X^2}{ (x-y)^2}\Big]
\end{eqnarray}
As shown in the appendix, the expression for $K_{2,1}$ can be simplified with the final result:
\begin{equation}
K_{2,1}(x,y,z)=\frac{\alpha_s^2N_c}{48 \pi}\frac{(x-y)^2}{X^2 Y^2}
\end{equation}

Note that eq.(\ref{k20}) determines $K_{2,0}$ in terms of $K_{3,0}$, but does not determine each coefficient function separately. Thus strictly speaking we need more information than is available to us directly to determine the virtual coefficients.
However, as we will show in the next sections, only a very specific expression for $K_{3,0}$ satisfies the condition of conformal invariance.
This form is:
\begin{equation}\label{k30}
K_{3,0}(w,x,y)\ =\ -\ \frac{1}{3}\left[ \int_{z,z'}K_{3,2}(w,x,y;z,z^\prime)\ +\ \int_z K_{3,1}(w,x,y;z)\right]\,.
\end{equation}
This expression is explicitly antisymmetric under the permutation of any two coordinates, and thus its action on dipole vanishes.
In the following we take the coefficient $K_{3,0}$ in (\cite{nlojimwlk}) to be given by eq.(\ref{k30}). Strictly speaking, this leaves a gap in our proof of conformal invariance which should be closed by explicit calculation of $K_{3,0}$ by a different method. However, given that we are able to show that conformal invariance does hold for all other terms, we consider this gap not to be significant.

With $K_{3,0}$ given by eq.(\ref{k30}), eq.(\ref{k20}) gives
\begin{eqnarray}\label{k20i}
K_{2,0}(x,y)&=&\frac{\alpha^2N_c}{16\pi^3}\int_z\frac{(x-y)^2}{X^2 Y^2}\Big[\frac{\pi^2}{3}+2 \ln\frac{Y^2}{ (x-y)^2}\,\ln\frac{X^2}{ (x-y)^2}\Big]
\end{eqnarray}
\section{Naive conformal transformations.}
It is now straightforward to find the conformal transformation properties of the Hamiltonian. 
The Hamiltonian is obviously rotationally and dilatationally invariant. It is therefore sufficient to consider the transformation of coordinate inversion. This is most conveniently done in the complex notation. For a 2d vector $x$, we introduce $x_{\pm}= x_1\pm i\,x_2$. The ``naive'' inversion transformation
is 
\beq {\cal I}_0:\  S(x_+,x_-)\rightarrow S(1/x_+,1/x_-) \ \ \ J_{L,R}(x_+,x_-)\rightarrow \frac{1}{x_+x_-}\,J_{L,R}(1/x_+,1/x_-)\,. \label{naiveconf}\eeq 

The transformation properties of the various kernels under the inversion are
\begin{eqnarray}
&& K_{3,2}(1/w;1/x,1/y;1/z,1/z^\prime)=z^4z^{\prime 4}\left[K_{3,2}(w;x,y;z,z^\prime)+\delta K_{3,2}(w;,x,y;z,z^\prime)\right], \ \ \ \ \nonumber\\
&&\delta K_{3,2}=\frac{i}{ 2}\Big[  M_{x,y,z}M_{y,z,z^\prime}+M_{x,w,z}M_{y,w,z^\prime} - M_{y,w,z^\prime} M_{x,z^\prime,z}-M_{x,w,z}M_{y,z,z^\prime}     \Big]\ln\frac{z^{\prime\,2}}{z^2}\\
\nonumber \\
&& K_{3,1}(1/w;1/x,1/y;1/z)= z^4\Big[ K_{3,1}(w;x,y;z)+ \delta K_{3,1}(w;x,y;z)\Big], \ \ \ \ \nonumber \\
&&\delta K_{3,1}(w;x,y;z)=\frac{i}{2}\int_{z^\prime}\Big[  \left (M_{x,w,z}-M_{y,w,z}\right)\left(M_{x,z,z^\prime}+M_{y,z,z^\prime}-M_{y,x,z^\prime}\right)\Big] \ln\frac{z^{\prime\,2}}{z^2}\\ \nonumber \\
&&K_{3,0}(1/w,1/x,1/y)=K_{3,0}(w,x,y)+\delta K_{3,0}(w,x,y)\nonumber\\
&&\delta K_{3,0}(w,x,y)=-\frac{i}{6}\int_{z,z'} z^4z'^4 \Bigg[ M_{x,w,z}M_{y,w,z'}+M_{x,y,z'}\left(M_{y,w,z}-M_{x,w,z}\right)+\nonumber \\
&&~~~~~~~~~~~~~~~~~~~~~~~~~~~~~~~~~~~~~~~~~~~+M_{x,w,z}M_{x,z,z'}-M_{y,w,z}M_{y,z,z'}\Bigg]\ln\frac{z^{\prime\,2}}{z^2}\\
&& K_{2,1}(1/x,1/y;1/z)= z^4K_{2,1}(x,y;z)\\ \nonumber \\
&&K_{2,0}(1/x,1/y)=K_{2,0}(x,y)\,.
\end{eqnarray}
The last line requires some explanation. Formally eq.(\ref{k20i}) gives
\begin{equation}\label{deltak20}
\delta K_{2,0}(x,y)=-iN_c\int_{z,z'}z^4z'^4\Big[-M_{x,y,z}M_{x,y,z'}+M_{x,y,z}M_{x,z,z'}+M_{x,y,z}M_{y,z,z'}\Big]\ln\frac{z^{\prime\,2}}{z^2}\,.
\end{equation}
However, the expression in the bracket is symmetric under $z\leftrightarrow z'$. This property is obvious for the first term, while it also holds for the other two terms since purely algebraically
\beq \label{identity}
M_{x,z,z^\prime}M_{x,y,z} =M_{x,y,z^\prime}M_{y,z^\prime,z}=\frac{\alpha_s^2}{ 4\pi^4}\ \frac{(x-y)^2}{(z-z^\prime)^2(y-z)^2(x-z^\prime)^2}
\eeq
The integrand in eq.(\ref{deltak20}) is therefore an antisymmetric function of $z$ and $z'$ and vanishes upon integration.
Thus the kernel $K_{2,0}$ is in fact conformally invariant.

For the very same reason we can disregard the last two terms in $\delta K_{3,0}$. 
The coefficients of these terms do not depend on one of the coordinates. As discussed above, such terms cannot be dropped automatically, but rather we should commute the charge associated with this coordinate to the right of all other operators before discarding a term of this type. However in the present case such commutation generates a term proportional to $\int_{z,z'}M_{x,y,z}M_{x,z,z'}\ln\frac{z^{\prime\,2}}{z^2}J^a(x)J^a(y)$. Since the operator $J^a(x)J^a(y)$ is symmetric under the interchange $x\rightarrow y$, the coefficient can be symmetrized and it vanishes due to eq.(\ref{identity}). We will therefore disregard these terms.

Under the action of the "naive" inversion the Hamiltonian transforms in the following way:
\begin{eqnarray}
&&{\cal I}_0\ H^{JIMWLK}\ {\cal I}_0\ =\ H^{JIMWLK}\,+\, {\cal A}; \\ \nonumber \\
&& {\cal A}=\int_{w,x,y, z,z^\prime}\delta K_{3,2}(w;x,y;z,z^\prime)f^{acb}\,\Big[J_L^d(x)\, J_L^e(y)\, S^{dc}_A(z)\,S^{eb}_A(z^\prime)\,J_R^a(w)\,\nonumber \\ &&-
 \,J_L^a(w)\,S^{cd}_A(z)\,S^{be}_A (z^\prime)\,J_R^d(x)\,J_R^e(y)\,\Big]+\nonumber \\
 &&+\int_{w,x,y, z}\, \delta K_{3,1}(w;x,y;z)\,f^{bde}\,\Big[  J_L^d(x) \,J_L^e(y) \,S^{ba}_A(z)\,J_R^a(w)\,-\, 
J_L^a(w)\,S^{ab}_A(z)\,J_R^d(x)\,J_R^e(y)\Big] \, +\, \nonumber \\
 &&+ \delta K_{3,0}(w;x,y)f^{bde}\,\Big[J_L^d(x) \,J_L^e(y) \,J_L^b(w)\,-\, 
J_R^c(x) \,J_R^b(y)\,J_R^a(w)\,\Big]\end{eqnarray} 
After substituting the above expressions for the $\delta K$s we obtain
\begin{eqnarray}
&& {\cal A}=\frac{i}{2}\int_{w,x,y, z,z^\prime}\ln\frac{z^{\prime\,2}}{z^2}\ \Bigg\{f^{acb}\Big[J_L^d(x)\, J_L^e(y)\, S^{dc}_A(z)\,S^{eb}_A(z^\prime)\,J_R^a(w)\nonumber \\ &&
- \,J_L^a(w)\,S^{cd}_A(z)\,S^{be}_A (z^\prime)\,J_R^d(x)\,J_R^e(y)\,\Big]\,\Big[M_{x,w,z}M_{y,w,z^\prime} - M_{y,w,z^\prime} M_{x,z^\prime,z}-M_{x,w,z}M_{y,z,z^\prime}     \Big]\nonumber\\
 &&+f^{bde}\,\Big[  J_L^d(x) \,J_L^e(y) \,S^{ba}_A(z)\,J_R^a(w)\,-\, 
J_L^a(w)\,S^{ab}_A(z)\,J_R^d(x)\,J_R^e(y)\Big]\nonumber\\
&&\times\Big[  \left (M_{y,w,z}-M_{x,w,z}\right)M_{y,x,z^\prime} +M_{x,w,z}M_{y,z,z^\prime}-M_{y,w,z}M_{x,z,z^\prime}\Big] -\frac{1}{3}f^{bde}\,\Big[J_L^d(x) J_L^e(y) J_L^b(w)
\nonumber\\
&&- 
J_R^d(x) J_R^e(y)J_R^b(w)\Big]\Big[ M_{x,w,z}M_{y,w,z'}+M_{x,y,z'}\left(M_{y,w,z}-M_{x,w,z}\right)\Big]\Bigg\}\nonumber\\
&&-\frac{N_c}{2}\int_{x,y,z,z'}\ln\frac{z^{\prime\,2}}{z^2}\ M_{x,y,z}\left(M_{x,z,z'}+M_{y,z,z'}\right)J_L^a(x)\,S_A^{ab}(z)\,J^b_R(y)\label{deltah}
%
\end{eqnarray}

In the first term we dropped the term in $\delta K_{3,2}$ which does not depend on $w$, since it vanishes when acting on color singlets. The last term ($JSJ$) arises from the term $-M_{y,w,z}M_{y,z,z^\prime}$ in $\delta K_{3,1}$ which does not depend on $x$. It is generated by commuting the appropriate global color charge $Q^a$ to the right of the rest of the factors in the operator $JJSJ$. As discussed above, once $Q^a$ has been commuted to the rightmost position it can be discarder. 

\section{The Conformal Symmetry of the Hamiltonian.}

 The NLO Hamiltonian is not invariant under the naive inversion transformation ${\cal I}_0$,
 \beq
 {\cal I}_0:\, H^{NLO\ JIMWLK}\rightarrow H^{NLO\ JIMWLK} \,+\, {\cal A}\,.
 \eeq
 One might however expect, that the Hamiltonian does possess an exact inversion (and conformal) symmetry, but that this symmetry is represented in a slightly different way than the naive transformation eq.(\ref{naiveconf}). This is generically the situation if one arrives at an effective theory by integrating out a subset of degrees of freedom. Say, one integrates over the subset $\{\alpha\}$ and obtains effective theory in terms of the remaining degrees of freedom $\{\beta\}$. If the cutoff separating $\alpha$ from $\beta$ is not invariant under a symmetry of the full theory, the transformation of $\beta$ involves $\alpha$, that is $\delta \beta=f(\alpha,\beta)$. After the integration $f(\alpha,\beta)$ becomes some effective operator expressible in terms of $\beta$ only. However this operator generically is not simply equal to $f(\alpha=0,\beta)$. This means, that the transformation of 
 $\beta$ in the effective theory looks somewhat different than in the original formulation before the integration of $\alpha$.
The situation in our case is very similar. The sharp rapidity cutoff used in deriving $H^{NLO\ JIMWLK}$ is not invariant under the conformal symmetry. Thus we expect that the naive form of conformal transformation should be modified, but that the symmetry itself is still the symmetry of $H^{NLO\ JIMWLK}$. 
  
If this is true, the anomalous piece ${\cal A}$ can be compensated if the Wilson lines $S$ form a non-trivial representation of the conformal group such that
\begin{eqnarray}
&&{\cal I}: S(x)\rightarrow S(1/x)\,+\, \delta S(x)\,,\ \  \ \ \ {\cal I}: J_{L,R}(x)\rightarrow \frac{1}{x^2}\,J_{L,R}(1/x)\,+\,  \delta J_{L,R}(x)\nonumber \\
&&{\cal I}: H^{LO} \rightarrow H^{LO}\,-\,{\cal A}
\end{eqnarray}
where $\delta S$ and $\delta J$ are perturbatively of the order $\alpha_s$, 
such that the net anomaly is cancelled and the total Hamiltonian remains invariant at NLO:
\beq
{\cal I}: H^{LO} \ +\  H^{NLO}\ \rightarrow \ H^{LO}\ +\ H^{NLO}\,.
\eeq
Note that within the NLO perturbative framework, the transformation of $H^{NLO}$ is with the ``naive'' operator ${\cal I}_0$ only.

Our goal is to explicitly construct such ${\cal I}$.  We are going to search it perturbatively in the form
\beq
{\cal I}\ =\ (1\ +\ {\cal C})\ {\cal I}_0\,.
\eeq
Inspired by the construction of the conformal dipole in \cite{N=4}, we search for $\cal C$  in the form:
\beq
{\cal C}\,=\,\int_{u,v,z^\prime} F(u,v,z^\prime)\,\hat h(u,v,z^\prime)\,,
\eeq
where $\hat h(u,v,z)$ is the leading order Hamiltonian density defined in eq.(\ref{LO}), and $F(u,v,z)$ is a function to be determined.

The result of the commutation (some details are given in Appendix C) is
\begin{eqnarray}\label{deltaLO}
&&\delta H^{LO}= -i\int_{x,y,w,z,z^\prime} f^{acb}\Big[S_{A}^{ec}(z)S_{A}^{db}(z^\prime)J_L^e(x) J_L^d(y) J_R^a(w)\ -\nonumber \\ 
&&~~~~~-S_{A}^{ce}(z)S_{A}^{bd}(z^\prime)J_R^e(x) J_R^d(y) J_L^a(w)\Big]
\Big[ F_{y,w,z^\prime}M_{w,x,z}-F_{w,x,z}M_{y,w,z^\prime}
-F_{y,w,z^\prime}M_{x,z^\prime,z}+\nonumber \\ 
&&~~~~~+F_{x,z^\prime,z}M_{y,w,z^\prime}- F_{y,z,z^\prime}M_{x,w,z}+ F_{x,w,z}M_{y,z,z^\prime}\Big]  \nonumber \\
&&+\int_{x,y,z,z^\prime} f^{abc} f^{def} S_{A}^{be}(z)S_{A}^{cf}(z^\prime)J_L^a(x)  J_R^d(y) \Big[ F_{x,y,z^\prime}M_{y,z^\prime,z}-F_{y,z^\prime,z}M_{x,y,z^\prime}-
\nonumber \\ 
&&~~~~~-F_{x,z,z^\prime}M_{y,x,z} +F_{y,x,z}M_{x,z,z^\prime}
\Big]  \nonumber \\
&&+i\, \int_{x,y,w,z,z^\prime} f^{bde}\Big[S_{A}^{ba}(z) J_L^d(x) J_L^e(y) J_R^a(w)\ -\ S_{A}^{ab}(z)J_R^d(x) J_R^e(y) J_L^a(w)\Big]\ \times \nonumber \\
&&~~~~~\times \Big[( F_{y,w,z} M_{y,x,z^\prime}-F_{x,w,z}M_{x,y,z^\prime}) +(F_{x,w,z}M_{z,y,z^\prime}-  F_{y,w,z} M_{x,z,z^\prime})  \nonumber\\
&&~~~~~+(F_{x,y,z^\prime}M_{x,w,z}-F_{y,x,z^\prime}M_{y,w,z}) + (F_{x,z,z^\prime}M_{y,w,z} -F_{z,y,z^\prime}M_{x,w,z}) \Big]\ + \nonumber \\
&&+\, N_c \int_{x,y,z,z^\prime} S_{A}^{ab}(z) J_L^a(x)  J_R^b(y)  \Big[ -F_{x,y,z} M_{z,x,z^\prime} +   M_{x,y,z} F_{z,x,z^\prime} -\nonumber \\
&&~~~~~ -F_{y,x,z} M_{z,y,z^\prime} +   M_{x,y,z} F_{z,y,z^\prime} \Big]\ + \nonumber \\
&&+\frac{i}{4}\int_{x,y,w,z,z^\prime}f^{bde}\left[J_R^d(x) J_R^e(y) J_R^b(w)-J_L^d(x) J_L^e(y) J_L^b(w)\right]\times\nonumber\\
&&~~~~~\times\Big[F_{x,y,z^\prime}(M_{x,w,z}-M_{y,w,z})-M_{x,y,z^\prime}(F_{x,w,z}-F_{y,w,z})+F_{y,w,z^\prime}M_{x,y,z}-M_{y,w,z^\prime}F_{x,y,z}\nonumber \\
&&~~~~~+M_{y,w,z^\prime}F_{x,w,z}-F_{y,w,z^\prime}M_{x,w,z}\Big]
\end{eqnarray}
Note that if $F(x,y,z)\,=\,M(x,y,z)\,\phi(z)$, the contribution to anomaly from the $SSJJ$ term vanishes due to eq.(\ref{identity}).

By inspection we see that the anomaly is cancelled in all terms if we choose
\beq F(x,y,z)\,=\,-\,\frac{1}{2}\,M(x,y,z)\,\ln \frac{z^2}{a^2}
\eeq
 with $a$ - an arbitrary constant. With this choice we have
\begin{eqnarray}
\delta H^{LO}&=& \frac{i}{2}\int_{x,y,w,z,z^\prime} \ln \frac{z'^2}{z^2}\ f^{acb}\Big[S_{A}^{ec}(z)S_{A}^{db}(z^\prime)J_L^e(x) J_L^d(y) J_R^a(w)\ -\nonumber \\
&-&S_{A}^{ce}(z)S_{A}^{bd}(z^\prime)J_R^e(x) J_R^d(y) J_L^a(w)\Big]\Big[ M_{y,w,z^\prime}M_{x,w,z}-
M_{y,w,z^\prime}M_{x,z^\prime,z}- M_{y,z,z^\prime}M_{x,w,z}\Big]  \nonumber \\
&-&\frac{i}{2}\, \int_{x,y,w,z,z^\prime} \ln \frac{z'^2}{z^2}\ f^{bde}\Big[S_{A}^{ba}(z) J_L^d(x) J_L^e(y) J_R^a(w)\ -\ S_{A}^{ab}(z)J_R^d(x) J_R^e(y) J_L^a(w)\Big]
\times \nonumber \\
&\times& \Big[ M_{x,y,z^\prime}(M_{x,w,z}-M_{y,w,z})+(M_{x,z,z^\prime}M_{y,w,z} -M_{y,z,z^\prime}M_{x,w,z})\Big]  \nonumber\\
&-&\, \frac{N_c}{2} \int_{x,y,z,z^\prime} \ln \frac{z'^2}{z^2}\ S_{A}^{ab}(z) J_L^a(x)  J_R^b(y)  \Big[  M_{x,y,z} (M_{x,z,z^\prime} + M_{y,z,z^\prime} )\Big]\ + \nonumber \\
&-&\frac{i}{8}\int_{x,y,w,z,z^\prime}\ln \frac{z'^2}{z^2}\ f^{bde}\left[J_R^d(x) J_R^e(y) J_R^b(w)-J_L^d(x) J_L^e(y) J_L^b(w)\right]\times\nonumber\\
&\times&\Big[M_{x,y,z^\prime}(M_{x,w,z}-M_{y,w,z})+M_{y,w,z^\prime}M_{x,y,z}-M _{y,w,z^\prime}M_{x,w,z}\Big]
\end{eqnarray}

We thus find that the Hamiltonian is indeed invariant under the transformation 
\beq\label{i}
{\cal I}\ S(x)\ {\cal I}= S(1/x) + [{\cal C},  S(1/x)] = S(1/x) -{\cal I}_0 [{\cal C},  S(x)]{\cal I}_0
\eeq
with 
\beq
{\cal C}\ =\ -\ \frac{1}{2}\ \int_{x,y,z}\,M(x,y,z)\,\ln \left(\frac{z^2}{a^2}\right)\ \hat h(x,y,z)\eeq
The last equality in eq.(\ref{i}) follows since
\beq\label{ioc}
{\cal I}_0\ {\cal C\ I}_0\, =\, -\, {\cal C}\,.
\eeq
The explicit operator form of the transformation can be read off eqs.(\ref{coms},\ref{coml},\ref{comr}). Note that the transformation of $S$ involves the operators $J_L$ and $J_R$. The simplest way to understand what is the result of this transformation, is rather than examining its operator form, directly examine its action on ``states''.
In other words we wish to examine its action on expectation values of operators in a state with vanishing $J_L$ and $J_R$. It is such expectation values that are
 the subject of the Balitsky hierarchy. Thus for example, acting on a dipole $[u^\dagger v]\equiv tr[S^\dagger(u)S(v)]$ we obtain
\begin{eqnarray}
{\cal I}[u^\dagger v]&=&{\cal I}_0[1-{\cal C}][u^\dagger v] \\
&=&\Bigg[[(1/u)^\dagger, 1/v]+\int_z M_{u,v,z}\ln\frac{z^2}{a^2}\left\{[(1/u)^\dagger, 1/z][(1/z)^\dagger, 1/v]-N_c[(1/u)^\dagger, 1/v]\right\}\Bigg]\nonumber \,.
\end{eqnarray}

\section{Constructing conformal operators.}

Our next step is to relate the modified conformal symmetry with the construction of the conformal dipole operator  in \cite{N=4}, and to extend such construction to arbitrary operators.
The original Wilson line operators transform in a non-canonical way under the modified conformal symmetry. One can, however construct operators which do transform in the standard way.

We define a conformal Wilson line operator $U(x)$ so that under the modified conformal (inversion) symmetry it transforms as
\beq
{\cal I}\ U(x)\ {\cal I}= U(1/x)
\eeq
It is easy to find a perturbative solution to this condition:
\beq
\bar U(x)\,=\,S(x)\,  +\,  \frac{1}{2}\,[{\cal C},  S(x)]\,.
\eeq
Indeed
\beq
{\cal I}\ \bar U(x) \ {\cal  I}=I  \Big(S(x)  +  \frac{1}{2}[{\cal C}, S(x)]\Big) I=S(1/x) + [{\cal C}, S(1/x)] - \frac{1}{2}[{\cal C},  S(1/x)]= \bar U(1/x)\,.
\eeq
The operator $\bar U$ has the requisite transformation properties under inversion, but it does not transform correctly under translation, since the kernel in operator ${\cal C}$ depends explicitly on the coordinate $z$ rather than on coordinate difference. This however is easily rectified.
Let us modify the definition 
\beq\label{invinv}
 U(x)\,=\,S(x)\,  +\,  \frac{1}{2}\,[\bar{\cal C},  S(x)]
\eeq
where $\bar {\cal C}={\cal C}-{\cal D}$, such that $D$ is invariant under inversion, but $\bar {\cal C}$ has correct transformation properties under translation. The choice of $\cal D$ is not unique, but a convenient choice that reproduces the results of \cite{N=4}  is
\beq
{\cal D}\,=\,-\,\frac{1}{2}\,\int_{x,y,z}\,M(x,y,z)\,\ln \frac{(x-y)^2z^2}{(x-z)^2(y-z)^2}\ \hat h(x,y,z)\,.
\eeq
This gives
\beq
\bar {\cal C}\,=\,\frac{1}{2}\,\int_{x,y,z}\,M(x,y,z)\,\ln \frac{(x-y)^2a^2}{(x-z)^2(y-z)^2}\ \hat h(x,y,z)\,.
\eeq
Similarly, for an arbitrary operator $O(x_1\cdots x_n)$ we can perturbatively define its conformal extension:
\beq
{\cal O}(x_1\cdots x_n)\,=\,O(x_1\cdots x_n)\,+\, \frac{1}{2}\,[\bar {\cal C},\,O(x_1\cdots x_n) ]\,.
\eeq
Applied to a single dipole this gives
\beq
tr[U^\dagger(u)U(v)]=[u^\dagger v]+\frac{1}{2}\int_z M(u,v,z)\ln \frac{(u-v)^2a^2}{(u-z)^2(v-z)^2}\left\{[u^\dagger z][ z^\dagger v]-N_c[u^\dagger v]\right\}
\eeq
which coincides with the conformal dipole of \cite{N=4}.

Another operator of interest is a three quark singlet operator  $$B(u,v,w)=\epsilon_{ijk}\epsilon_{lmn}\,S^{im}_F(u)S^{jl}_F(v)S^{kn}_F(w)$$ for $N_c=3$. Its conformal extension is calculated to be
\begin{eqnarray}
&&{\cal B}(u,v,w)=B(u,v,w)+\frac{3}{4}\int_z \left\{ M_{u,v,z}\ln\frac{(u-v)^2a^2}{(u-z)^2(v-z)^2} \Big[\frac{1}{ 6}(B(u,z,z)B(w,v,z)+\right. \nonumber \\ &&+B(v,z,z)B(w,u,z)-B(w,z,z)B(v,u,z)) 
 -\left. B(u,v,w)\Big]+(u\leftrightarrow w)+(v\leftrightarrow w)\right\}\,. \nonumber \\
\end{eqnarray}
Linearized NLO evolution equation for ${\cal B}$ should coincide with the result of ref. \cite{NLOBKP}, but this comparison is beyond the scope of the present paper.

Finally, we note that one can rewrite the JIMWLK Hamiltonian in terms of the conformal Wilson line operators $U(x)$. This is achieved by a substitution,
 inverting eq.(\ref{invinv})
\beq
 S(x)\,=\,U(x)  \,-  \,\frac{1}{2}\,[\bar {\cal C},  U(x)]
\eeq
or alternatively by transforming $H^{JIMWLK}$ with the operator $\bar {\cal C}$
\beq
{\cal H}^{JIMWLK}_{conformal}\,=\,H^{JIMWLK}\,-\,\frac{1}{2}[\bar {\cal C},  H^{LO\ JIMWLK}]\,.
\eeq
The calculation of this commutator is entirely analogous to eq. (\ref{deltaLO}) with a different function $F$.
The "conformal" Hamiltonian ${\cal H}^{JIMWLK}_{conformal}$ has exactly the same structure as the original  NLO JIMWLK Hamiltonian  (\ref{NLO})
but with the kernels $K$ replaced by conformal kernels $\cal K$:
\begin{eqnarray}\label{k1}
&& {\cal K}_{3,2}(w;x,y;z,z^\prime)
 =\frac{i}{ 4}\Big[ M_{x,y,z}M_{y,z,z^\prime}\ln\frac{W^4X^{\prime\,2}Y^{\prime\,2}}{W^{\prime\,4}X^2Y^2}
 +M_{x,w,z}M_{y,w,z^\prime} \ln\frac{(x-w)^2 W^2Y^{\prime\,2}}{(y-w)^2W^{\prime\,2}X^2} \nonumber \\
&& - M_{y,w,z^\prime} M_{x,z^\prime,z}\ln\frac{ W^4 X^{\prime\,2}Y^{\prime\,2}}{(y-w)^2 (z-z^\prime)^2W^{\prime\,2}X^2}
 -M_{x,w,z}M_{y,z,z^\prime}  \ln\frac{ (x-w)^2 (z-z^\prime)^2 W^2 Y^{\prime\,2}}{W^{\prime\,4}X^2Y^2}
    \Big]; \nonumber\\ \nonumber \\
 && {\cal K}_{3,1}(w;x,y;z)=  \int_{z^\prime}\, \left[{\cal K}_{3,2}(y;w,x;z,z^\prime)\,-\,{\cal K}_{3,2}(x;w,y;z,z^\prime)\right];
\nonumber \\ \nonumber \\
  && {\cal K}_{2,2}(x,y;z,z^\prime)= {K}_{2,2}(x,y;z,z^\prime)+\nonumber \\
  &&+ \frac{\alpha_s^2}{16\pi^4}\frac{(x-y)^2}{(z-z^\prime)^2}
  \left[ \frac{1}{Y^2X^{\prime\,2}} \ln \frac{(x-y)^2(z-z^\prime)^2}{X^2Y^{\prime\,2}} 
  + \frac{1}{X^2Y^{\prime\,2}} \ln \frac{(x-y)^2(z-z^\prime)^2}{Y^2X^{\prime\,2}} \right];
  \nonumber \\ \nonumber \\
  && {\cal K}_{2,1}(x,y;z)= {K}_{2,1}(x,y;z)-\nonumber \\
  &&- \frac{\alpha_s^2\,N_c}{16\pi^4}\int_{z^\prime} \frac{(x-y)^2}{(z-z^\prime)^2}
  \left[ \frac{1}{Y^2X^{\prime\,2}} \ln \frac{(x-y)^2(z-z^\prime)^2}{X^2Y^{\prime\,2}} 
  + \frac{1}{X^2Y^{\prime\,2}} \ln \frac{(x-y)^2(z-z^\prime)^2}{Y^2X^{\prime\,2}} \right]; \\ 
  \nonumber  \\
  &&
{\cal K}_{3,0}(w,x,y)\ =\ -\ \frac{1}{3}\left[ \int_{z,z'}{\cal K}_{3,2}(w,x,y;z,z^\prime)\ +\ \int_z {\cal K}_{3,1}(w,x,y;z)\right];\\ \nonumber  \\
&& {\cal K}_{2,0}(x,y,z)\ =\  {K}_{2,1}(x,y;z).
\end{eqnarray}
For a dipole, ${\cal H}^{JIMWLK}_{conformal}$ generates an evolution equation that fully agrees with the evolution of conformal dipole
given by eq. (66) of \cite{N=4}.

\section{Conclusions}
In this paper we studied conformal symmetry of the NLO JIMWLK Hamiltonian in ${\cal N}=4$ theory. We showed that even though the Hamiltonian was derived using 
a sharp rapidity cutoff, which is not conformally invariant, conformal symmetry indeed remains an exact symmetry of $H^{NLO\ JIMWLK}$.
The action of the conformal transformation on the Wilson line operators acquires an additional term, consistent with the fact that those are operators in the effective theory obtained by integrating out part of the degrees of freedom.

We also showed how to define conformal extension for any function of Wilson lines so that the resulting operator has the standard ``naive'' conformal symmetry 
transformation properties.   As two examples we considered the dipole operator and the baryon operator in the $SU(3)$ theory. We have also provided the expression for 
$H^{NLO\ JIMWLK}$ expressed in terms of the ``conformal'' Wilson lines.

The  conformal extension can be applied to $H^{NLO\ JIMWLK}_{QCD}$ presented in eq. (\ref{NLO1}).  While the resulting Hamiltonian
is not conformal,  the conformal invariance of this new Hamiltonian is only broken by terms proportional to the QCD $\beta$ function.

\appendix
\section{ QCD NLO JIMWLK Hamiltonian}

Here we quote the main result  of \cite{nlojimwlk} for NLO JIMWLK Hamiltonian in QCD.
\begin{eqnarray}\label{NLO1}
&&H^{NLO\ JIMWLK}_{QCD}= \int_{x,y,z}K_{JSJ}(x,y;z) \left[ J^a_L(x)J^a_L(y)+J_R^a(x)J_R^a(y)-2J_L^a(x)S_A^{ab}(z)J^b_R(y)\right] + \nonumber \\
 &&+\int_{x\,y\, z\,z^\prime} K_{JSSJ}(x,y;z,z^\prime)\left[f^{abc}f^{def}J_L^a(x) S^{be}_A(z)S^{cf}_A(z^\prime)J_R^d(y)- N_c J_L^a(x)S^{ab}_A(z)J^b_R(y)\right] 
 +\nonumber \\
 &&+\int_{x,y, z,z^\prime} K_{q\bar q}(x,y;z,z^\prime)\left[2\,J_L^a(x) \,tr[S^\dagger(z)\, T^a\,S (z^\prime)T^b]\,J_R^b(y)\,
 -\, J_L^a(x)\,S^{ab}_A(z)\,J^b_R(y)\right]+\nonumber \\
 &&+\int_{w,x,y, z,z^\prime}K_{JJSSJ}(w;x,y;z,z^\prime)f^{acb}\,\Big[J_L^d(x)\, J_L^e(y)\, S^{dc}_A(z)\,S^{eb}_A(z^\prime)\,J_R^a(w)\,-\nonumber \\
 &&-J_L^a(w)\,S^{cd}_A(z)\,S^{be}_A (z^\prime)\,J_R^d(x)\,J_R^e(y)\,+\frac{1}{3}[\,J_L^c(x) \,J_L^b(y) \,J_L^a(w)\,-\, 
J_R^c(x) \,J_R^b(y)\,J_R^a(w)]\,\Big] \,
 +\nonumber \\
 &&+\int_{w,x,y, z}\, K_{JJSJ}(w;x,y;z)\,f^{bde}\,\Big[  J_L^d(x) \,J_L^e(y) \,S^{ba}_A(z)\,J_R^a(w)\,-\, 
J_L^a(w)\,S^{ab}_A(z)\,J_R^d(x)\,J_R^e(y) \, +\, \nonumber \\
 &&+\frac{1}{3}[\,J_L^d(x) \,J_L^e(y) \,J_L^b(w)\,-\, 
J_R^d(x)\,J_R^e(y)\,J_R^b(w)]
\Big] 
\end{eqnarray}
All $J$s in  (\ref{NLO}) are assumed not to act on $S$ in the Hamiltonian.  
\begin{eqnarray}
&&K_{JSJ}(x,y;z) =-\frac{\alpha_s^2}{16 \pi^3}
\frac{(x-y)^2}{X^2 Y^2}\Big[b\ln(x-y)^2\mu^2
-b\frac{X^2-Y^2}{ (x-y)^2}\ln\frac{X^2}{Y^2}+
(\frac{67}{9}-\frac{\pi^2}{ 3})N_c-\frac{10}{ 9}n_f\Big]\nonumber \\
&&- \frac{N_c}{2}\ \int_{z^\prime}\, \tilde K(x,y,z,z^\prime)
\end{eqnarray}
Here $\mu$ is the normalization point in the $\overline{MS}$ scheme and
$b=\frac{11}{3}N_c-\frac{2}{3}n_f$. \begin{eqnarray}
&&K_{JSSJ}(x,y;z,z^\prime) \ =\ \frac{\alpha_s^2}{16\,\pi^4}
\Bigg[\,-\,\frac{4}{ Z^4}\,+\,
\Big\{2\frac{X^2{Y'}^2+{X'}^2Y^2-4(x-y)^2Z^2}{ Z^4[X^2{Y'}^2-{X'}^2Y^2]}\nonumber\\ 
&&+
~\frac{(x-y)^4}{ X^2{Y'}^2-{X'}^2Y^2}\Big[
\frac{1}{X^2{Y'}^2}+\frac{1}{ Y^2{X'}^2}\Big]
+\frac{(x-y)^2}{Z^2}\Big[\frac{1}{X^2{Y'}^2}-\frac{1}{ {X'}^2Y^2}\Big]\Big\}
\ln\frac{X^2{Y'}^2}{ {X'}^2Y^2}\Bigg]\,+\nonumber \\
&&+\tilde K(x,y,z,z^\prime) \end{eqnarray}
\begin{eqnarray}
&&\tilde K(x,y,z,z^\prime)=\frac{i}{2}\,\left[K_{JJSSJ}(x;x,y;z,z^\prime)-K_{JJSSJ}(y;x,y;z,z^\prime)-\right.\nonumber \\
&&~~~~~~~~~~~~~~~~~~~~~~~~~~~~~~~~~~~\left. -K_{JJSSJ}(x;y,x;z,z^\prime)+K_{JJSSJ}(y;y,x;z,z^\prime)\right]
\end{eqnarray}
\begin{eqnarray}
K_{q\bar q}(x,y;z,z^\prime) =-\frac{\alpha_s^2\,n_f}{ 8\,\pi^4}
\Big\{
\frac{{X'}^2Y^2+{Y'}^2X^2-(x-y)^2Z^2}{ Z^4(X^2{Y'}^2-{X'}^2Y^2)}
\ln\frac{X^2{Y'}^2}{ {X'}^2Y^2}-\frac{2}{Z^4}\Big\}
\end{eqnarray}
\begin{eqnarray}
&&K_{JJSJ}(w;x,y;z)\,=\,-\,i\,\frac{\alpha_s^2}{ 4\, \pi^3 }\,\Big[ \frac{X\cdot W}{ X^2\,W^2}\,-\, \frac{Y\cdot W}{ Y^2\,W^2}    \Big] \ln\frac{Y^2}{ (x-y)^2}\,\ln\frac{X^2}{ (x-y)^2}\\
&&K_{JJSSJ}(w;x,y;z,z^\prime)=-i
\frac{\alpha_s^2}{ 2\,\pi^4}
\left(\frac{X_iY^\prime_j}{ X^2Y^{\prime 2}}
\right)\Big(\frac{\delta_{ij}}{2 Z^2}-\frac{Z_i W^\prime_j}{ Z^2 W^{\prime 2}}+
\frac{Z_j W_i}{ Z^2 W^{ 2}}-\frac{W_i W^\prime_j}{W^2 W^{\prime 2}}
\Big)\ln\frac{W^2}{ {W'}^2} \nonumber \\ 
\end{eqnarray}

\section{Technical details of derivations}
\subsection{Action of the NLO JIMWLK Hamiltonian on the dipole}

To facilitate comparison with the results of \cite{BC} we present the action of various operators on the dipole $[u^\dagger v]$
\begin{eqnarray}\label{kaction}
&&\int_{x,y,z}K_{2,1}(x,y,z)J_L^a(x)S_A^{ab}(z)J^b_R(y)[u^\dagger v]=\int_z K_{2,1}(u,v,z)\left\{\frac{1}{N_c}[u^\dagger v]-[u^\dagger z][z^\dagger v]\right\}\\
&& \int_{x,y}\, K_{2,0}(x,y)\, \left[ J^a_L(x)\,J^a_L(y)\,+\,J_R^a(x)\,J_R^a(y)\right][u^\dagger v]=-4\int_z K_{2,0}(u,v,z)\frac{N^2_c-1}{2N_c}[u^\dagger v]\\
&&\int_{w,x,y, z}K_{3,1}(w;x,y;z)f^{bde}\Big[  J_L^d(x) J_L^e(y) S^{ba}_A(z)J_R^a(w)-
J_L^a(w)S^{ab}_A(z)J_R^d(x)J_R^e(y) \Big][u^\dagger v]=\nonumber\\\
&&~~~~~~~~~~~~=-iN_c\int_z\Bigg[K_{3,1}(v;u,v;z)+K_{3,1}(u;v,u;z)\Bigg]\left\{[u^\dagger z][z^\dagger v]-\frac{1}{N_c}[u^\dagger v]\right\}\\
&&\int_{x,y, z,z^\prime}\, K_{2,2}(x,y;z,z^\prime)f^{abc}\,f^{def}\,J_L^a(x) \,S^{be}_A(z)\,S^{cf}_A(z^\prime)\,J_R^d(y)[u^\dagger v]\nonumber \\
&&~~~~~~~~~~~~~~~~~~~~~~~~~~
=-\int_{z,z'}K_{2,2}(u,v;z,z^\prime)\left\{[u^\dagger z'][z'^\dagger z][z^\dagger v]-[u^\dagger zz'^\dagger v z^\dagger z']\right\} \\
&&\int_{w,x,y, z,z^\prime}K_{3,2}(w;x,y;z,z^\prime)f^{acb}\Big[J_L^d(x) \,J_L^e(y)\, S^{dc}_A(z)\,S^{eb}_A(z^\prime)\,J_R^a(w)\,-\nonumber \\
&&~~~~~~~~~~~~~~~~~~~~~~~~~~~~~~~~~~~~~~~~~~~ -  \,J_L^a(w)\,S^{cd}_A(z)\,S^{be}_A (z^\prime)\,J_R^d(x)\,J_R^e(y)\,\Big][u^\dagger v]=\nonumber\\
 &&~~~~~~~=\frac{i}{2}\int_{z,z^\prime}\Bigg\{\Big[K_{3,2}(v;v,u;z,z')+K_{3,2}(u;u,v;z,z')-K_{3,2}(u;v,u;z,z')-\nonumber \\
 && ~~~~~~~-K_{3,2}(v;u,v;z,z')\Big]  \left\{[u^\dagger z'][z'^\dagger z][z^\dagger v]-[u^\dagger zz'^\dagger v z^\dagger z']\right\}\ +\ 
\Big[K_{3,2}(v;v,u;z,z')-\nonumber \\
&&~~~~~~~-K_{3,2}(u;u,v;z,z')-K_{3,2}(u;v,u;z,z')+K_{3,2}(v;u,v;z,z')-K_{3,2}(v;u,u;z,z')+\nonumber\\
 &&~~~~~~~~~~~~~~~~~~~~~~~~~~~~~~~~~~~~~~~~~+
K_{3,2}(u;v,v;z,z')\Big]\ [u^\dagger z'][z'^\dagger z][z^\dagger v]\Bigg\}\\
 &&\int_{w,x,y} K_{3,0}(w,x,y)f^{bde}\left[\,J_L^d(x) \,J_L^e(y) \,J_L^b(w)\,-\, 
J_R^d(x)\,J_R^e(y)\,J_R^b(w)\right][u^\dagger v]=\nonumber\\
&&~~~~~~~~=i\frac{N^2_c-1}{2}\Big\{K_{3,0}(v,u,v)+K_{3,0}(u,v,u)-K_{3,0}(v,v,u)-\nonumber \\
&&~~~~~~~~~~~~~~~~~~~~~~~~~~~~~~~~~~-K_{3,0}(u,u,v)+K_{3,0}(v,u,u)+K_{3,0}(u,v,v)\Big\}[u^\dagger v]
\end{eqnarray}
We note, that with the explicit expression for $K_{3,2}$ and $K_{3,1}$ in eq.(\ref{k}), we have
\begin{eqnarray}
&&K_{3,2}(v;v,u;z,z')+K_{3,2}(u;u,v;z,z')-K_{3,2}(u;v,u;z,z')-K_{3,2}(v;u,v;z,z')=0\nonumber\\ \nonumber \\
&&K_{3,1}(v;u,v;z)+K_{3,1}(u;v,u;z)=-\frac{i}{2}\int_{z'}\Bigg[M_{u,v,z}M_{u,v,z'}\ln \frac{U^2V^2}{U'^2V'^2}+
\nonumber \\ 
&&+ M_{u,v,z'}(M_{v,z,z'}-M_{u,z,z'})\ln \frac{U'^2V^2}{U^2V'^2}-2M_{u,v,z}\Big[M_{u,z,z'}\ln\frac{V^2}{V'^2}+M_{v,z,z'}\ln\frac{U^2}{U'^2}\Big]\Bigg]
\end{eqnarray}
\subsection{Algebra with kernels.}
We can relate the kernels $K_{3,2}$ and $K_{3,1}$ to the kernels  $K_{JJSSJ}$ and $K_{JJSJ}$ introduced above.
\begin{eqnarray}
&&\frac{\alpha_s^2N_c}{8 \pi^3}\frac{(x-y)^2}{X^2 Y^2} \ln\frac{Y^2}{ (x-y)^2}\,\ln\frac{X^2}{ (x-y)^2}=-\frac{i}{2}N_c\Big[K_{JJSJ}(x;x,y;z)-K_{JJSJ}(y;x,y;z)\Big]=\nonumber\\
&&=-\frac{i}{2}N_c\int_{z'}\Big[K_{JJSSJ}(y;x,x;z,z')-K_{JJSSJ}(y;y,x;z,z')+K_{JJSSJ}(x;y,y;z,z')-\nonumber \\
&&~~~~~~~~~~~~~~~~~~~~~~~~~~~~~~~~~~~~~~~~~~~~~~~~~~~~~~~~~~~~~~~~~~~~~~~~~ -K_{JJSSJ}(x;x,y;z,z')\Big]
\end{eqnarray}
We can relate $K_{JJSSJ}$ with $K_{3,2}$ by straightforward algebraic manipulations:
\begin{eqnarray}
&&K_{JJSSJ}(w;x,y,z,z')=K_{3,2}(w;x,y,z,z')-\nonumber\\
&&-i\frac{\alpha_s^2}{8\pi^4}\Bigg[\frac{2\pi^2}{\alpha_s}\Big[-M_{y,z,z'}\frac{1}{W^2}-M_{x,z',z}\frac{1}{W'^2}+M_{y,w,z'}\left(\frac{1}{W^2}-\frac{1}{Z^2}\right)+M_{x,w,z}\left(\frac{1}{W'^2}-\frac{1}{Z^2}\right)\Big]\nonumber\\
&&-\frac{1}{W'^2}\frac{1}{W^2}+\frac{1}{Z^2}\left(\frac{1}{X^2}+\frac{1}{Y'^2}+\frac{1}{W^2}+\frac{1}{W'^2}\right)\Bigg]\ln\frac{W^2}{W^{\prime\,2}} 
\end{eqnarray}
This allows us to express $K_{2,1}$ as
\begin{eqnarray}
&&K_{2,1}(x,y,z)=\frac{\alpha_s^2N_c}{48 \pi}\frac{(x-y)^2}{X^2 Y^2}-\\
&&-\frac{\alpha_s^2N}{16\pi^4}\int_{z'}\Bigg\{\frac{2\pi^2}{\alpha_s}\Bigg[(M_{y,z',z}-M_{x,z',z})\left[\frac{1}{Y'^2}\ln\frac{Y^2}{Y'^2}-\frac{1}{X'^2}\ln\frac{X^2}{X'^2}\right]+
\nonumber \\
&&+M_{x,y,z}\left[\frac{1}{Y'^2}\ln\frac{Y^2}{Y'^2}+\frac{1}{X'^2}\ln\frac{X^2}{X'^2} -\frac{1}{Z^2}\ln\frac{X^2Y^2}{X'^2Y'^2}\right]\Bigg]
+\frac{1}{Z^2}\left(\frac{1}{X^2}-\frac{1}{Y^2}\right)\ln\frac{X'^2Y^2}{X^2Y'^2}\Bigg\}\nonumber
\end{eqnarray}
The last term in this equation can be simplified if we discard the terms that do not depend on either $x$ or $y$. With some additional algebra we have
\begin{equation}
K_{2,1}(x,y,z)=\frac{\alpha_s^2N_c}{48 \pi}\left(1-\frac{6a}{\pi^3}\right)\frac{(x-y)^2}{X^2 Y^2}
\end{equation}
where the constant $a$ is defined by
\begin{equation}
\int_{Y'}\frac{Y'_i}{Y'^2(Y-Y')^2}\ln\frac{Y^2}{Y'^2}=\xi\frac{Y_i}{Y^2}
\end{equation}
The last equation must be true by rotational invariance and dimensional counting, given that the integral is convergent.
The explicit calculation gives $\xi=0$. Thus we determine the coefficient $K_{2,1}$ as
\begin{equation}
K_{2,1}(x,y,z)=\frac{\alpha_s^2N_c}{48 \pi}\frac{(x-y)^2}{X^2 Y^2}
\end{equation}

\section{Transforming the leading order Hamiltonian}
Using the basic commutation relations:
\begin{eqnarray}
&&[J_R^a(x),J_R^b(y)]=if^{abc}J^c_R(x)\delta(x-y)\,,\ 
[J_L^a(x),J_L^b(y)]=-if^{abc}J^c_L(x)\delta(x-y)\,, \\
&&[J_R^a(x),S_A^{bc}(y)]=if^{acd}S_A^{bd}(x)\delta(x-y)\,,\ 
[J_L^a(x),S_A^{bc}(y)]=-if^{abd}S_A^{dc}(x)\delta(x-y)\nonumber
\end{eqnarray}
we obtain
\begin{eqnarray}\label{coms}
\delta S_{A}^{ab}(z)&\equiv&[S_{A}^{ab}(z),{\cal C}]\\
&=& i\,\int_{v,z^\prime} F_{z,v,z^\prime}\Big[f^{cad}[S_{A}^{db}(z)J_L^c(v)-2S_{A}^{db}(z)S_{A}^{ce}(z^\prime)J_R^e(v)]-f^{cbd} S_{A}^{ad}(z)J_R^c(v)\Big]
\nonumber \\
&-& i\,\int_{u,z^\prime} F_{u,z,z^\prime}\Big[f^{cbd}[J_R^c(u)S_{A}^{ad}(z)-2J_L^e(u)S_{A}^{ad}(z)S_{A}^{ec}(z^\prime)]
-f^{cad} J_L^c(u) S_{A}^{db}(z^\prime)\Big]\nonumber
\end{eqnarray}
\begin{eqnarray}\label{coml}
\delta J_L^a(x)&\equiv&[J_L^a(x),{\cal C}]=- if^{abc}\,\int_{v,z^\prime} F_{x,v,z^\prime}\Big[J_L^c(x)J_L^b(v)-2J_L^c(x)S_{A}^{bd}(z^\prime)J_R^d(v)\Big]-\nonumber \\
&-& if^{abc}\int_{u,z^\prime} F_{u,x,z^\prime}J_L^b(u)J_L^{c}(x)
+2if^{abd} \int_{u,v} F_{u,v,x}J_L^b(u) S_A^{dc}(x)J_R^c(v)
\end{eqnarray}
\begin{eqnarray}\label{comr}
\delta J_R^a(x)&\equiv&[J_R^a(x),{\cal C}]= if^{abc}\,\int_{u,z^\prime} F_{u,x,z^\prime}\Big[J_R^b(u)J_R^c(x)-2J_L^d(u)S_{A}^{db}(z^\prime)J_R^c(x)\Big]\nonumber \\
&+& if^{abc}\int_{v,z^\prime} F_{x,v,z^\prime}J_R^c(x)J_R^{b}(v)
-2if^{acd} \int_{u,v} F_{u,v,x}J_L^b(u) S_A^{bd}(x)J_R^c(v)
\end{eqnarray}
The transformation of the LO JIMWLK Hamiltonian
\begin{eqnarray}
&&\delta H^{LO}\equiv[H^{LO}, {\cal C}]= -\frac{1}{2}\int_{x,y,z}M(x,y,z)\times \\
&&\times\Big\{J_R^a(y)[J_R^a(x),{\cal C}]+[J_R^a(x),{\cal C}] J_R^a(y)+
J_L^a(y)[J_L^a(x),{\cal C}]  +[J_L^a(x),{\cal C}] J_L^a(y) -\nonumber \\ &&
-2[J_L^a(x),{\cal C}] S_{A}^{ab}(z)J_R^b(y)- 2J_L^a(x)[S_{A}^{ab}(z),{\cal C}]J_R^b(y)-2J_L^b(y)S_{A}^{ba}(z)[J_R^a(x),{\cal C}] 
\Big\}\nonumber 
\end{eqnarray}
Putting all terms together yields the transformation of $H^{LO}$, eq.(\ref{deltaLO}).

\section*{Acknowledgments}
We are most grateful to Ian Balitsky who encouraged us to think about this project. 
We thank S. Caron Huot for interesting correspondence. M.L and Y.M. thank the Physics Department of the University of Connecticut for hospitality at the time when
this project was initiated.
The research was supported by the DOE grant DE-FG02-92ER40716; the EU FP7 grant PIRG-GA-2009-256313; the  ISRAELI SCIENCE FOUNDATION grant \#87277111;  the EU FP7 IRSES network "High-Energy QCD for Heavy Ions";  and the BSF grant \#2012124.


\begin{thebibliography}{99}

\bibitem{GLR}  L.V.~Gribov, E.~Levin and M.~Ryskin, Phys. Rep.
100:1,1983.

\bibitem{cgc}   E.Iancu, A. Leonidov and L. McLerran, {\it Nucl. Phys.} 
{\bf A 692} (2001) 583; {Phys. Lett.} {\bf B
510} (2001) 133;
E. Ferreiro, E. Iancu, A. Leonidov, L. McLerran;  
{\it Nucl. Phys.}{\bf A703} (2002) 489.
             
\bibitem{jimwlk} J. Jalilian Marian, A. Kovner, A.Leonidov and H.
Weigert,
{\it Nucl. Phys.}{\bf  B504} 415 (1997); 
{\it Phys. Rev.} {\bf D59} 014014 (1999); 
J. Jalilian Marian, A. Kovner and H. Weigert, {\it Phys. Rev.}{\bf D59} 
014015 (1999); 
A. Kovner and J.G. Milhano, {\it Phys. Rev.} {\bf D61} 014012 (2000) .
 A. Kovner, J.G. Milhano and H. Weigert,
{\it Phys.Rev.} {\bf D62} 114005 (2000); 
 H. Weigert, {\it Nucl.Phys.} {\bf A 703} (2002) 823;
 
 
\bibitem{Bal}
I. Balitsky, {\it Nucl. Phys.}  {\bf B463} 99 (1996); 
{\it Phys. Rev. Lett.} {\bf 81} 2024 (1998); 
{\it Phys. Rev.}{\bf D60} 014020 (1999).




\bibitem{bfkl}
 V.~S.~Fadin, E.~A.~Kuraev and L.~N.~Lipatov,
  Phys.\ Lett.\  B {\bf 60} (1975) 50;
  Sov. Phys. JETP
                {\bf 45} (1977) 199 ; \\
          Ya. Ya. Balitsky and L. N. Lipatov,
               {  Sov. J. Nucl. Phys.}\, {\bf 28} (1978) 22.
               
              
\bibitem{bkp}
J.~Bartels,
%
Nucl.\ Phys.\  {\bf B175}, 365 (1980);\,\,\,\,
J.~Kwiecinski and M.~Praszalowicz,
%
Phys.\ Lett.\  {\bf B94}, 413 (1980);

\bibitem{KOV}
  Y.~V.~Kovchegov,
  Phys.\ Rev.\ D {\bf 61}, 074018 (2000)
  [arXiv:hep-ph/9905214].
  CITATION = HEP-PH 9905214;

\bibitem{lublinsky} E. Gotsman, E. Levin, M. Lublinsky and  U. Maor; Eur.Phys.J. C27 (2003) 411-425; e-Print: hep-ph/0209074

\bibitem{AAMQS} J. L. Albacete, N. Armesto, J. G. Milhano, P. Quiroga-Arias and C. A. Salgado; Eur.Phys.J. C71 (2011) 1705;
e-Print: arXiv:1012.4408 [hep-ph] 

\bibitem{review} J. L. Albacete ,  A. Dumitru and C. Marquet; Int.J.Mod.Phys. A28 (2013) 1340010; [arXiv:1302.6433 [hep-ph]]
 
\bibitem{NLOBFKL} V. S. Fadin  and L.N. Lipatov, Phys.Lett.B429:127-134,1998.
e-Print: hep-ph/9802290;
G.Camici and M. Ciafaloni, Phys. Lett. B 430 (1998) 349.

\bibitem{BalNLO} 
  I.~Balitsky,
  Phys.\ Rev.\ D {\bf 75}, 014001 (2007)
  [hep-ph/0609105];
 
               
  \bibitem{weigertrun} Y.~V.~Kovchegov and H.~Weigert,
  Nucl.\ Phys.\  A {\bf 784} (2007) 188
  [arXiv:hep-ph/0609090];
  Nucl.\ Phys.\ A {\bf 789}, 260 (2007)
  [hep-ph/0612071];
  E.~Gardi, J.~Kuokkanen, K.~Rummukainen and H.~Weigert,
  Nucl.\ Phys.\ A {\bf 784}, 282 (2007)
  [hep-ph/0609087].
  
\bibitem{beuf} G. Beuf, J.Phys.Conf.Ser. 422 (2013) 012026; e-Print: arXiv:1301.0773 [hep-ph] 
  




\bibitem{BC} 
  I.~Balitsky and G.~A.~Chirilli,
  Phys.\ Rev.\ D {\bf 77}, 014019 (2008)
  [arXiv:0710.4330 [hep-ph]].

\bibitem{N=4} I. Balitsky and G. Chirilli, Nucl. Phys. B822 (2009) 45-87; e-Print: arXiv:0903.5326 [hep-ph].


  
\bibitem{Grab} 
  A.~V.~Grabovsky,
  JHEP {\bf 1309}, 141 (2013)
  [arXiv:1307.5414 [hep-ph]].

\bibitem{nlojimwlk} A. Kovner, M. Lublinsky and Y. Mulian;  arXiv:1310.0378 [hep-ph]

\bibitem{BClast}   I.~Balitsky and G.~A.~Chirilli, "Rapidity evolution of Wilson lines at the next-to-leading order", Phys.\ Rev.\ D {\bf 88}, 111501 (2013);
e-Print: arXiv:1309.7644 [hep-ph].

\bibitem{Simon} S. Caron-Huot, "The next-to-leading order Balitsky-JIMWLK equation", to appear.


\bibitem{KLMatwork} A. Kovner, M. Lublinsky and Y. Mulian, in preparation.


\bibitem{FF}  V.S. Fadin, R. Fiore, Phys. Lett. B661, 139 (2008).


\bibitem{reggeon} A. Kovner and M. Lublinsky, JHEP 0702:058,2007;
e-Print: hep-ph/0512316.


\bibitem{iancutri} E. Iancu and  D.N. Triantafyllopoulos, JHEP 1204 (2012) 025; e-Print: arXiv:1112.1104 [hep-ph].



  
  
  
  
                
 \bibitem{NLOBKP} 
  J.~Bartels, V.~S.~Fadin, L.~N.~Lipatov and G.~P.~Vacca,
  Nucl.\ Phys.\ B {\bf 867}, 827 (2013)
  [arXiv:1210.0797 [hep-ph]].
  
  







  

 













    


  

  
  
  
  
   

                    


            
               
               
               






  
 
 
 





 


  

  


\end{thebibliography}
\end{document}